\documentclass[aps,prb,twocolumn,groupedaddress,showpacs,showkeys,floatfix]{revtex4-1}
\usepackage{amsmath}
\usepackage{amsfonts}
\usepackage{amssymb}
\usepackage{graphicx}

\begin{document}

%\preprint{}

\title{Graphyne on metallic surfaces: an improved graphene}
\author{P.~Lazi\' c$^1$ and \v Z.~Crljen$^{1,2}$}

\affiliation{
$^1$Theoretical Physics Division, R. Bo\v skovi\' c Institute, Zagreb, Croatia\\
             $^2$Faculty od Sciences, University of Split, Split, Croatia}

\date{\today}

\begin{abstract}

We showed how a structural modification of graphene, which gives a carbon 
allotrope graphyne, can induce an energy gap at the K point of the Brillouin 
zone. Upon adsorption on metallic surfaces, the same mechanism is responsible
for the further modification of the energy gap which occurs via the charge 
transfer mechanism. We performed the calculation based on the density 
functional theory with the novel non-local vdW-DF correlation of the 
adsorption of graphyne on Cu(111), Ni(111) and Co(0001) surfaces and showed 
the dependence of the gap change on the charge transfer in the system. 
The binding of graphyne appears to be stronger than of graphene on the 
same surfaces.

\end{abstract}

\maketitle

\section{Introduction}

Carbon nanostructures, and graphene in particular, are becoming unavoidable 
materials in the growing filed of nanoelectronics. The experimental 
realization\cite{Nov,Nov2} of a single 
layer graphene has boosted a tremendous interest in its physical properties.
The most prominent feature of this two-dimensional material is its exceptional 
electronic structure with linear band dispersion in the vicinity of the Dirac 
point. Graphene shows a huge mobility of charge carriers, high 
conductivity of electrons and holes and a ballistic charge transport, what 
makes it a promising candidate for a number of applications in nanoscale 
electronic devices.

However, what seems to be a graphene strongest side is also its weakest point.
Namely, graphene band structure has no band gap what makes it unusable in 
building some of basic electronic elements. Graphene field-effect-transistor, 
for example, cannot be turned off effectively due to the absence of a bandgap.
Creating a bandgap in graphene is one of the most important 
research topics in graphene community since it may ultimately enable new 
applications in electronics, nanospintronics, and infrared
nanophotonics. A number of approaches have been proposed or implemented to 
create a bandgap in graphene already, such as using graphene-substrate 
interaction\cite{GR-subs}, lateral confinement\cite{lat-conf}, uniaxial 
strain\cite{strain} and breaking the inversion symmetry in bilayer 
graphene\cite{bilay1,bilay2}.

Here, an alternative approach is proposed, the use of another allotrope of 
carbon instead, a graphyne. Our presentation is organized as follows. 
The motivation for our studies is further supported in section II where 
the freestanding graphyne is discussed. In section III we report for the 
first time the graphyne adsorption on three representative metallic
surfaces showing the features of binding of graphyne on Co(0001), 
Ni(111) and Cu(111). Finally, in section IV we give conclusions and 
perspectives of graphyne.

\section{Free standing graphyne}
\begin{figure}[htbp]
%\resizebox{1.14\columnwidth}{!}{
\includegraphics[clip=true,width=0.95\columnwidth]{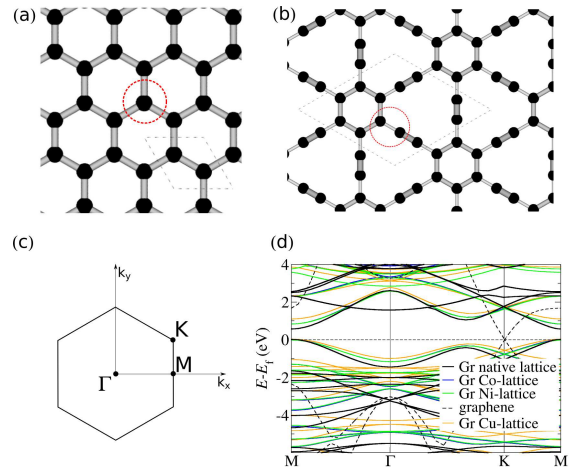}
%}
\caption{(Color online). Structure and energy bands. (a) Graphene, 
(b) Graphyne, (c) Reciprocal lattice,
(d) Energy bands for different lattice constant (see text). Circles in (a) 
and (b) denote the substitution in graphene that creates graphyne.}
\label{graphyne}
\end{figure}

Graphyne, shown in Fig.\ref{graphyne}(b), is a result of a structural change 
in the two dimensional graphene plane which opens a gap at the K point, i.e. 
lifts the degeneracy caused by the symmetry of the hexagonal structure. 
It can be easily understood by making a comparison with the one-dimensional 
linear polymer chain consisting of strongly interacting coplanar $p_{z}$ 
orbitals, each of which contributes one electron to the
resultant continuous $\pi$-electron system. The chain should behave 
essentially as a one-dimensional metal with a half-filled conduction 
band and show Peierls distortion and a metal-insulator transition. 
By introducing a bond alternation 
(alternating single long and double short bonds) and consequently 
doubling the lattice constant such a chain can efficiently lower its energy. 
\begin{figure*}[htp]
\centering
\includegraphics[clip=true,width=0.85\linewidth]{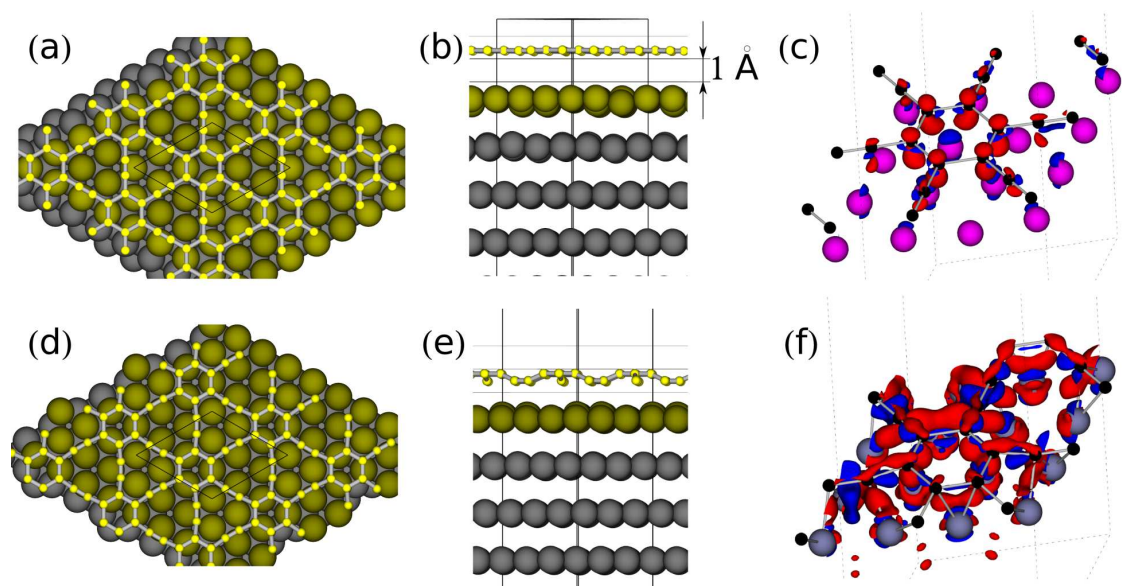}
\caption{(Color online). Graphyne adsorbed on Cu(111) in top row and Ni(111) 
in bottom row, (a) and (d) top view, (b) and (e) side view, (c) and 
(f) charge rearrangement (isosurface of 0.04 e/$\mathrm{\AA^{3}}$ plotted,
depletion in blue, accumulation in red), angle view, only the top layer
of surface atoms are plotted. 
}
\label{charges}
\end{figure*}
This reduces the extent of electronic delocalization that can take place 
along the backbone but opens an energy gap in the electronic structure of
the polymer. In a graphene each carbon atom, too, has one out of plane
$p_{z}$ orbital with one electron. The peculiarity of graphene, dictated
by the backbone hexagonal structure, is that each atom is surrounded with 
three equivalent neighbors  and cannot make such a transition with bond 
alternation. 
The periodic crystal potential is such that it causes a band splitting at
the $\Gamma$ and M symmetry points but degeneracy (zero bandgap) at 
the K points of the first Brillouin zone making graphene 
a zero bandgap semiconductor as shown in Fig.\ref{graphyne}(d). 

One way to lift that degeneracy is then to introduce an inequivalent neighbor 
to the hexagonal atom. In a  $\pi$ conjugated system this is easy to achieve 
by forcing a single bonding to the neighboring atom, as discussed for the 
case of a linear polymer. In Fig.\ref{graphyne}(b) it is achieved by 
introducing an acetylene in the hexagonal structure. The $p_{z}$ orbitals 
of the acetylene atoms are involved in a triple bond and acetylene atom 
can only form a single bond with the hexagonal atom. The consequence of the 
formation of a single bond is thus an inequivalent neighbor, the change of the 
periodicity of the system and lifting of the degeneracy at the K points.
The resulting graphyne has a high electron density at 
the Fermi level like the graphene, but most importantly it 
has a well defined direct bandgap of 1.1 eV at the M point of the Brillouin 
zone as displayed in Fig.\ref{graphyne}(d).

Graphyne and similar non-natural carbon allotropes  can be assumed to be 
chemically stable\cite{GDY-polar}, 
in spite of having a triple bond in their structure. Finite building 
blocks have  being synthesized\cite{synth1,synth2} and steps towards 
a preparation have been developed\cite{prepar1,synth1}. 
In fact,
a graphdiyne, an allotrope with diacetylenic linking chain, has already 
being synthesized\cite{GRD-film, GRD-wire} 
but only as a multi-layer film and a nanowire with the diameter of 30 nm. 
There is no report of a free standing or supported monolayer material.

The calculation of the band structure of single layer graphyne and 
graphdiyne have been reported\cite{GY1,GY2,GY3,GRribbon},
but to the best of our knowledge the adsorbed monolayer on any surface
has not yet been investigated.

\section{Graphyne on metallic surfaces}
Support of surfaces and coupling to metal contacts  play a fundamental 
role in technological applications
and may bring a new property resulting in novel devices.
In spintronics a hybrid structure of graphene and ferromagnetic surfaces,
for example, brings a promise for a spin-filtering device\cite{spintr1}.
In that respect graphyne as a semiconductor with different charge 
distribution locally is a new element, possibly a new building block 
in nanoelectronics.

The change it brings can be best verified investigating its adsorption
on transition metal surfaces.
We choose Cu(111), Ni(111) and Co(0001) surfaces, studied a lot in the 
context of graphene adsorption\cite{adsorp1,adsorp2,adsorp3}, 
as they provide the possibility for physisorption and chemisorption 
due to a hybridization of carbon orbitals with metallic states. 
Graphyne is expected to have a large polarizability\cite{GDR-polar}, 
which together with the polarizability of metallic surfaces may result
in a significant van der Waals interaction.
%\subsection{Calculations}
To calculate the structure and electronic properties we use the state of 
the art approach of density functional theory (DFT) as 
implemented in the VASP 5.1. computer code\cite{vasp} and vdW-DF 
correlation\cite{vdW-DF,nonlocal} with the optB88 
exchange\cite{adsorp3}. To describe the bonding, from physisorption to 
chemisorption, accurately we used a
recent selfconsistent implementation \cite{klimes1, klimes2} of the 
nonlocal vdW-DF functional  following the method
of R\'{o}man-P\'{e}rez and Soler\cite{PerSol}.
We employed a plane wave cutoff of 500 eV and dipole correction.
All structures were allowed to relax until the atomic forces were below 
1 meV/$\mathrm{\AA}$. The metallic slab was simulated by five 
layers of atoms, all of which were allowed to relax, and an additional
22  $\mathrm{\AA}$ of vacuum to avoid periodic image interaction. We have 
enlarged the graphyne lattice constant (a = 6.95 $\mathrm{\AA}$) for
several percents in order to make its unit cell commensurate 
with the (3x3) cell of the substrate metal\cite{binding}. 
It resulted in minor changes of band energies, Fig.\ref{graphyne}(d).

Four high symmetry positions of adsorbed graphyne were calculated: 
H1 and H2, TOP and BRIDGE. In H1 and H2 the center of grapyne's hexagonal 
ring was positioned above the FCC and HCP hollow sites of the unit cell, 
respectively.
For all three substrates the H1 and H2 proved to be the best adsorption
sites, as shown in Table~\ref{E-d}.
The total energies for H1 and H2 structures on Cu(111) surface were 
virtually identical, the H1 being just slightly lower. For Ni(111) and 
Co(0001) surfaces, H2 appeared to be energetically favored.

\begin{table}[b]
\caption{\label{E-d}
Graphyne (Gy) adsorbed on Cu, Ni and Co surfaces:
energetically favored adsorption sites, adsorption energy per carbon atom 
and the length of the triple bond (acetylenic link).
}
\begin{ruledtabular}
\begin{tabular}{cccc}
  & Adsorp. site & $E_{ads}$ / C atom  & $d_{CC3}$  \\ \hline
      \cline{2-4}
      free graphyne &   &   & 1.225 $\mathrm{\AA}$\\
      Gy/Cu(111) & H1  & 145 meV & 1.309 $\mathrm{\AA}$ \\
      Gy/Ni(111) & H2  & 351 meV & 1.383 $\mathrm{\AA}$\\
      Gy/Co(0001) & H2  & 376 meV & 1.401 $\mathrm{\AA}$
\end{tabular}
\end{ruledtabular}
\end{table}

%\subsection{Results}
Recalling that the values of adsorption energy per C atom for graphene 
on the Cu(111) surface is 38 - 68 meV\cite{graphene/Cu},
graphyne with 145 meV, as seen in Table~\ref{E-d}, evidently binds more 
strongly. It is a consequence of a different distribution of charges locally
in a graphyne plane. One expects a stronger interaction with Cu atoms 
located under the acetylenic link, then under the carbon ring.
But even though the interaction is different locally 
the graphyne plane does not show any corrugation, as shown in the top panel 
of Fig.\ref{charges}(b). The corrugation is shown in the substrate Cu plane 
instead where the Cu atoms located under the 
acetylenic link are slightly pulled out of plane. 
It is quite the opposite compared to the adsorption on Ni and Co surfaces.
Adsorption energies per carbon atom on those surfaces, 351 meV and 376 meV
respectively, are also larger compared 
to those of graphene (67 meV for Ni(111)\cite{adsorp3,co1} and 77 meV for
Co(0001)\cite{co1, co2} surfaces).
but Ni and Co surfaces are harder and do not relax as much as Cu surface. 
Their interaction with graphyne is evidently stronger compared to Cu as
their average distance to graphyne plane is considerably shorter.
Graphyne on Ni(111), Fig.\ref{charges}(b), and on Co(0001) (not shown here),
therefore suffers a strong corrugation, 
what is an indication of an inequivalent bonding to surfaces locally.

The extent of the variation of interaction can be better understood by
looking at the charge transfer in the system presented in Fig.\ref{charges}(c).
From the shapes of the charge depletion (blue) and charge
accumulation (red) regions, we conclude that there are several types of 
charge transfer. There is some accumulation of charges along the line 
between the acetylinic carbon and hexagonal carbon atom.
That have a consequence on the interaction in the graphyne backbone  
itself, as we have discussed before, leading to the change of the gap at 
the K point of the Brillouin zone.
There is a charge transfer between acetylenic carbon and the underlying 
substrate atoms, too, with the  accumulation of charges into 
the carbon $p_{z}$ orbital coming mostly from the $d$-orbital of the 
underlying substrate atom. 
In addition, the polarization along the graphyne plane as well as of 
the substrate atoms takes place which is a clear fingerprint of a 
vdW-bonded $\pi$-conjugated system. 
All of these effects are more pronounced for graphyne on Ni(111), 
Fig.\ref{charges}(c) bottom, 
than on Cu(111) surface, Fig.\ref{charges}(c) top. 
Surprisingly, on Co(0001) the charge transfer (not shown here) is very 
similar to that on Ni(111) surface although the surfaces are different.
The graphyne adsorption is governed by the vdW interaction
with a typical nonlocal correlation interaction features. In addition, 
a strong orbital hybridization of graphyne orbitals with the states of 
metal occurs, consistent with a small distances
of less then 2 $\mathrm{\AA}$, Fig.\ref{charges}(b). 

The calculated energy bands 
show a strong change compared to free standing graphyne bands.
In order to follow the details of the electronic structure changes in 
the figures the contribution of C $p_{z}$ states is made proportional 
to the thickness of the line. 

\begin{figure}[htbp]
\includegraphics[clip=true,width=0.85\columnwidth]{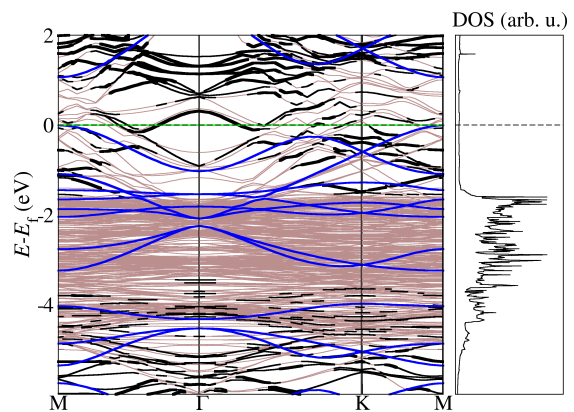}
\caption{(Color online). Energy bands of graphyne/Cu(111) (left): the 
contribution of Cpz orbital projection marked in black,
the free graphyne bands are in blue and density of states (DOS) of 
clean Cu (right).}
\label{Cu-bands}
\end{figure}

\begin{figure*}[htp]
\includegraphics[clip=true,width=0.42\linewidth]
{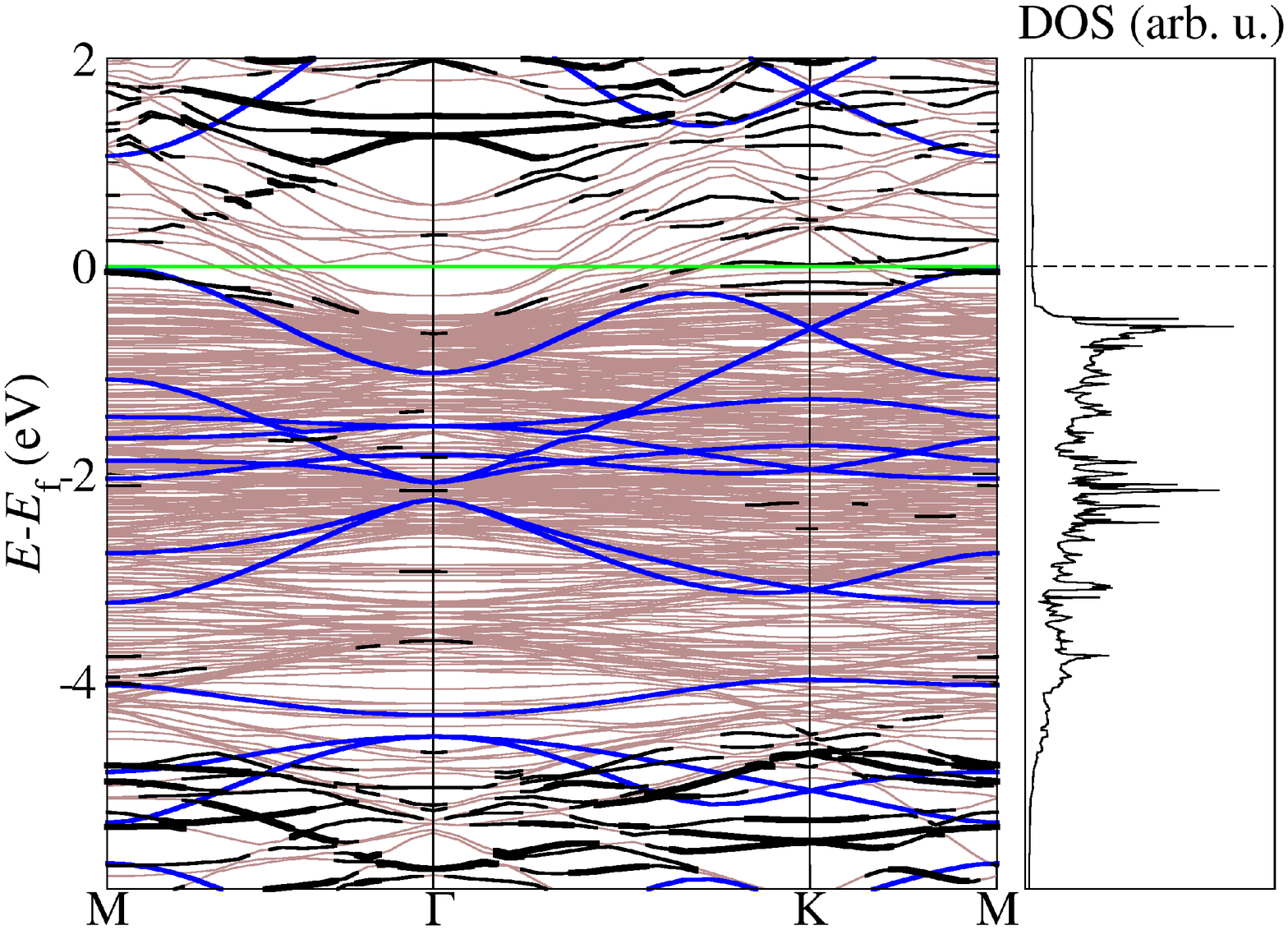}\hspace{0.4in}
\includegraphics[clip=true,width=0.42\linewidth]{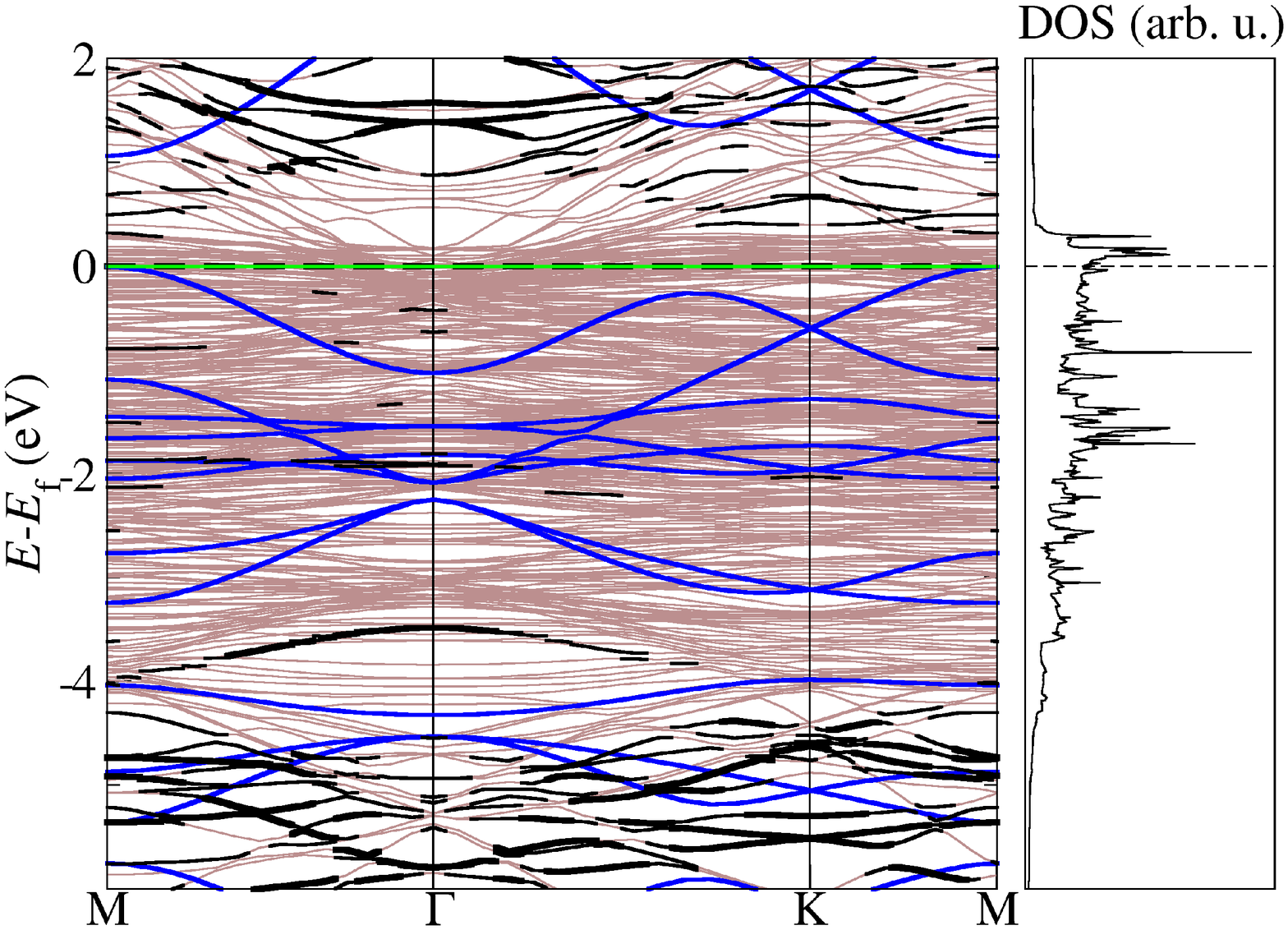}
\caption{(Color online). Energy bands of graphyne/Ni(111) and density of 
states of clean Ni for majority (left) and minority spin (right). 
The thickness of black lines show projection weight on carbon atoms. 
Blue lines - free standing graphyne.}
\label{Ni,bands}
\end{figure*}

\begin{figure*}[htp]
\includegraphics[clip=true,width=0.42\linewidth]
{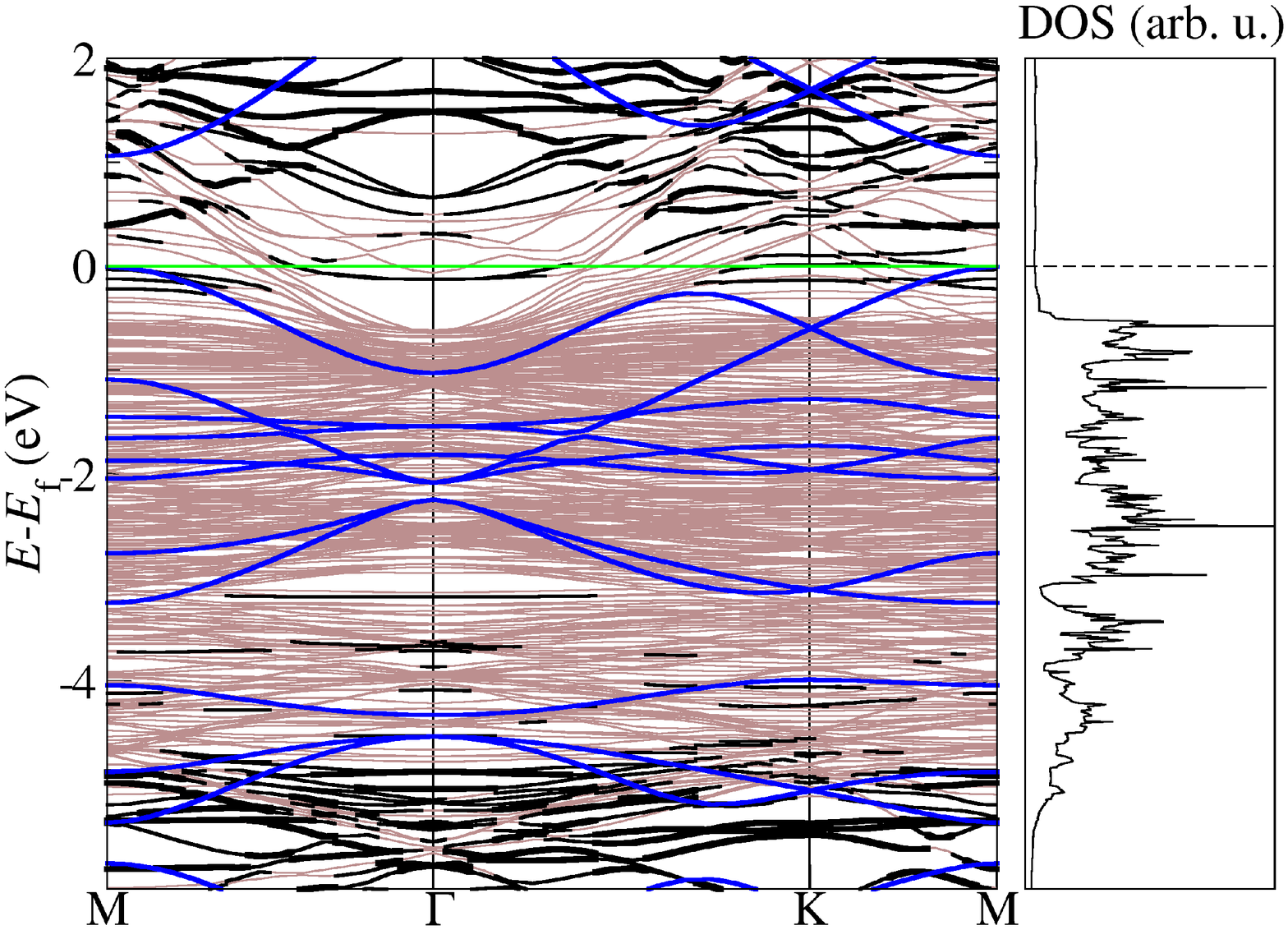}\hspace{0.4in}
\includegraphics[clip=true,width=0.42\linewidth]{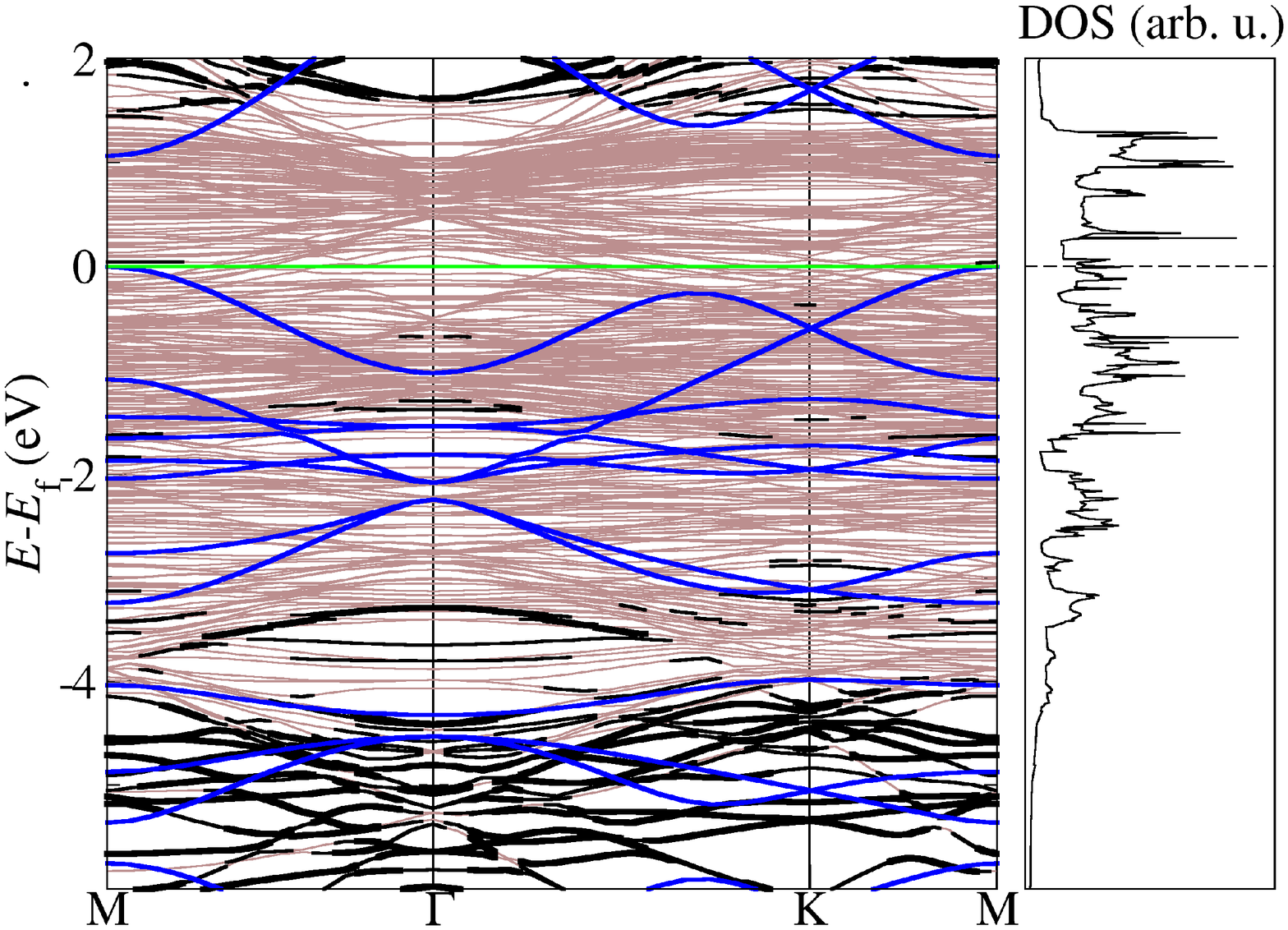}
\caption{(Color online). Energy bands of graphyne/Co(0001) and density of 
states of clean Co for majority (left) and minority spin (right). 
The thickness of black lines show projection weight on carbon atoms. 
Blue lines - free standing graphyne.}
\label{Co,bands}
\end{figure*}
As shown in Fig.\ref{Cu-bands} for graphyne on Cu the hybridization is 
particularly strong in the range of the metal d-band energy region, where 
the graphyne states are virtually dispersed over the band. 
Comparing the thicker band lines with the lines of free standing graphyne
we conclude on relative downshift of graphyne's levels and crossing at
Fermi level which renders graphyne metallic rather then semiconducting. 

On Ni(111) surface due to its ferromagnetic property  
different electronic structures are induced for majority and minority
spin states as seen in Fig.\ref{Ni,bands}. The overall downshift of graphyne's levels for both spins 
appears, similar to the Cu(111) surface. Due to the fact that the  $d$-band 
straddles the Fermi level the level repulsion of hybridized and unoccupied 
graphyne states gives no crossing at the Fermi level and the majority spin 
states around the Fermi level are mostly metallic $d$-states. 
There are pronounced gaps at K and M points of the Brillouin zone.
On the ferromagnetic Co(0001) surface, shown in Fig.\ref{Co,bands}, the band 
structures are very similar to those on Ni(111) surface taking the fact
that the d-band of cobalt is wider.

One notices that the length of a triple carbon bond, given in 
Table~\ref{E-d}, is longer in adsorbed then in free standing graphyne
(d = 1.225 $\mathrm{\AA}$)
and increasing from 1.309 $\mathrm{\AA}$ for Cu to 1.383 $\mathrm{\AA}$ 
for Ni and 1.401 $\mathrm{\AA}$ for Co surfaces. 
The increase is accompanied with the accumulation of charges between
acetylenic and hexagonal carbon atom as mentioned before. According to 
our discussion in section II, it implies the change of the bonding with 
hexagonal carbon tending to a more symmetric bonding between hexagonal 
atoms and its neighbors and consequently should be followed by the reduction 
of energy gaps at the K points. 
Indeed, the inspection of the structure of energy bands in 
Fig. 3., 4. and 5. supports the conclusion. The structure is 
partly blurred by the hybridization induced rather strong redistribution 
of energy levels in the vicinity of Fermi level, 
but apparently the energy gap is the most reduced at the K points and 
the reduction of the gap is increasing from Cu(111) to Ni(111) and 
Co(0001) surfaces. 

\section{Conclusions}
We have shown how a structural modification of graphene induces an energy 
gap and that the gap of the new allotrope graphyne can be modulated via 
the charge transfer upon adsorption on different metallic surfaces. As a
semiconductor with the density of states comparable to that of graphene,
a graphyne can become a valuable material in the field of nanoelectronics. 
We have calculated the adsorption of graphyne on Cu, Ni and Co surfaces
using a DFT method with a selfconsistent implementation of vdW-DF functional 
which explicitly includes nonlocal correlations. The binding of graphyne
appears to be  several times stronger than of graphene on the same
surfaces and a way above typical thermal fluctuations at room
temperature ($\sim25meV$), what makes a graphyne less prone to desorption. 
In conclusion our results show that a graphyne can complement and even 
overcome the graphene as a building block in future 
nanoelectronic components.

\begin{acknowledgments}
This work was supported in part by the Croatian Ministry of Science 
under Contract No. 098-0352828-3118. 
\end{acknowledgments}
%\bibliography{basename of .bib file

\begin{thebibliography}{00}
\bibitem{GDR-polar} Graphene is known to have large polarizability. 
The conductivity of graphdiyne is $2.516\times10^{-4} S m^{-1}$, 
and its large polarizability has been reported in Ref.\onlinecite{GDY-polar}.
\bibitem{binding} Adjusting the lattice constant of adsorbate or substrate 
is a common practice in calculation of graphene adsorption. It
significantly reduces the computational cost, but for some substrates 
the binding may depend strongly on the choice, see Ref. \onlinecite{co1}.
In reality graphene compensates over large distances with a moire pattern.
Graphyne is more prone to chemisorption and large distance rearrangements 
are not probable.
\end{thebibliography}

\end{document}